# High-performance graphene photodetector by interfacial gating

Xitao Guo,[1,∥] Wenhui Wang,[1,∥] Haiyan Nan,[1] Yuanfang Yu,[1] Jie Jiang,[1] Weiwei Zhao,[2] Jinhuan Li,[1] Zainab Zafar,[3] Nan Xiang,[2] Zhonghua Ni,[2] Weida Hu,[4] Yumeng You,[3,*] and Zhenhua Ni[1,*]

[1]Department of Physics and Key Laboratory of MEMS of the Ministry of Education, Southeast University, Nanjing 211189, China.

[2]Jiangsu Key Laboratory for Design and Fabrication of Micro-Nano Biomedical Instruments, School of mechanical engineering, Southeast University, Nanjing 211189, China.

[3]Ordered Matter Science Research Center, Southeast University, Nanjing 211189, China.

[4]National Laboratory for Infrared Physics, Shanghai Institute of Technical Physics, Chinese Academy of Sciences, 500 Yutian Road, Shanghai 200083, China.

[∥]These authors contributed equally to this work.

[*]Corresponding authors: zhni@seu.edu.cn, youyumeng@seu.edu.cn

**Graphene based photo-detecting has received great attentions and the performance of such detector is stretching to both ends of high sensitivity and ultra-fast response. However, limited by the current photo-gating mechanism, the price for achieving ultra-high sensitivity is sacrificing the response time. Detecting weak signal within short response time is crucial especially in applications such as optical positioning, remote sensing, and biomedical imaging. In this work, we bridge the gap between ultra-fast response and ultra-high sensitivity by employing a graphene/SiO$_2$/lightly-doped-Si architecture with revolutionary interfacial**

**gating mechanism. Such device is capable to detect < 1 nW signal (with responsivity of ~1000 A $W^{-1}$) and the spectral response extends from visible to near-infrared. More importantly, the photoresponse time of our device has been pushed to ~400 ns. The current device structure does not need complicated fabrication process and is fully compatible with the silicon technology. This work will not only open up a route to graphene-based high performance optoelectronic devices, but also have great potential in ultra-fast weak signal detection.**

## 1. INTRODUCTION

Graphene-based photodetectors have aroused considerable interest and various types of device configurations and mechanisms have been developed [1-8]. Current prototype devices have shown outstanding performance with individual functionalities aiming for different applications, that is, ultra-fast or ultra-sensitive detection. On the fast-detecting side, benefited from the high mobility and ultrafast carrier dynamics, intrinsic graphene based photodiode has shown photoresponse at a fs timescale [2]. On the ultra-sensitive side, by employing the photo-gating mechanism, hybrid graphene photoconductor has exhibited ultra-high gain up to $10^{10}$ [8]. However, there is a huge gap between the two mechanisms, like two sides of a coin: a fs detection only has a responsivity of ~mA $W^{-1}$ and a pW detection responses only in milliseconds to seconds timescale, while numerous applications such as optical positioning, remote sensing, biomedical imaging, desire both speed and sensitivity. The gap between the binary performances is limited by the current mechanisms employed. A fast detecting relies on the high carrier mobility of intrinsic graphene and suffers from its gapless nature and low efficiency of electron-hole pair disassociation. While the bottleneck of photo-gating is the slow charge transfer and/or charge trapping process in the time scale of ~ms, or even seconds [3,5,7-17], which is indeed necessary for the charges in the channel to recirculate between source and drain, to give rise to ultra-high gain.

In this work, by adopting a new concept of interfacial gating effect from lightly-doped silicon(Si)/silicon dioxide ($SiO_2$) interface, we successfully bridge the gap between ultra-fast response and ultra-sensitivity of graphene based photodetector. Such device architecture separates the photoexcited electron-hole pairs by intrinsic self-built electric field at $Si/SiO_2$ interface, and in turn the accumulated charges at the interface would gate graphene and introduce high gain of photoresponse by taking advantage of the high mobility of graphene. This charge transfer free strategy with fast accumulation of photoexcited carrier at the interface ensures the fast response of photocurrent at the graphene channel. Moreover, the current device structure does not need any complicated fabrication process and is fully compatible with the silicon technology.

## 2. DEVICE FABRICATION AND METHODS

### A. Device fabrication

Monolayer graphene samples are mechanically exfoliated from highly oriented pyrolytic graphite, and deposited on lightly p-doped Si substrate that is terminated with 300 nm of $SiO_2$. Source and drain electrodes (5 nm Ni adhesion layer, followed by a 50 nm Au capping layer) are defined using electron beam lithography (FEI, FP2031/12 INSPECT F50) and deposited by thermal evaporation (TPRE-Z20-IV). More than ten devices are fabricated and all of them show very good photoresponse behaviour. In addition, other control devices on 300 nm $SiO_2$/heavily-doped Si, and lightly-doped Si covered by $SiO_2$ with different thicknesses or 30 nm $Al_2O_3$ are also fabricated. The $Al_2O_3$ layer is deposited by using the atomic layer deposition (Sunaletmr-100).

### B. Photoresponse measurement

Electrical and photoresponse characteristics of the devices are measured using a Keithley 2612 analyzer under dark and illuminated conditions. Light is switched on and off by using an optical chopper or acoustic optical modulator (R21080-1DS) at different frequencies. The light source is an $Ar^+$ laser with wavelength of 514 nm. A super continuum light source (SuperK Compact ns kHz) is employed to attain the spectral photocurrent response. In all the photocurrent measurements, the laser and super continuum light are focused on the sample with a 50x objective (NA= 0.5) and the spot size of light is ~1 μm, much smaller than the graphene channel length. In the power dependent experiment, optical attenuators are introduced to change the input power. A digital storage oscilloscope (Tektronix TDS 1012, 100 MHz/1GS/s) is used to measure the transient response of photocurrent.

## 3. RESULTS AND DISCUSSION

### A. Characterization of graphene photodetector

Figure 1a shows the schematic diagram and a representative optical microscopy image of our device. A lightly p-doped silicon wafer (1-10 Ω cm) and thermally-grown 300 nm thick $SiO_2$ layer are employed as the gate electrode and dielectric, respectively. We have compared the Si substrates with different doping concentrations and the above mentioned one provides the best device performance. The mechanically exfoliated monolayer graphene is characterized by optical contrast and Raman spectroscopy (see Supplement 1, Figure S1a) [18]. The G band and 2D band (with a full width at half maximum of 27.7 $cm^{-1}$) locate at 1580.3 and 2675.2 $cm^{-1}$, respectively, and the ratio of $I_{2D}/I_G$ is ~2.3. Figure S1b shows a transfer characteristic of the device at an applied bias voltage $V_D = 10$ mV measured in dark, suggests that graphene is slightly p-doped because of interactions with the substrates, and the absorbed water/oxygen molecules in air. The estimated holes mobility (μ) is ~5,000 $cm^2$ $V^{-1}$ $s^{-1}$. All these features indicate the high quality of monolayer graphene and device.

### B. Mechanism of photodetector by interfacial gating effect

The working principle of our graphene photodetector can be understood through the energy band diagram of oxide-silicon interface and its effect on graphene as shown in Figure 1b and 1c. The localized interface states such as positive charge states ($q\varphi_0$) with energies within the silicon bandgap exist at the oxide-silicon interface, induce a negative depletion layer (-) in silicon near the interface and the formation of built-in electric field (E) [19]. Due to the presence of built-in electric field, the photogenerated electron-hole pairs in lightly p-doped Si will be separated: the holes (red points) in the valence band of Si diffuse toward the bulk Si, while the electrons (blue points) accumulate at the $SiO_2$/Si interface (Figure 1b). This leads to the appearance of a negative voltage at the interface, which can be negligible in heavily doped silicon due to the very short lifetime of the photogenerated carriers [20]. As a result, the additional negative voltage could effectively gate the graphene channel through capacitive coupling, lowers the Fermi level ($E_{f(Gr)}$) to its new position ($E'_{f(Gr)}$), as shown in Figure 1c. Therefore, the increase of hole density and high positive photocurrents in graphene are achieved.

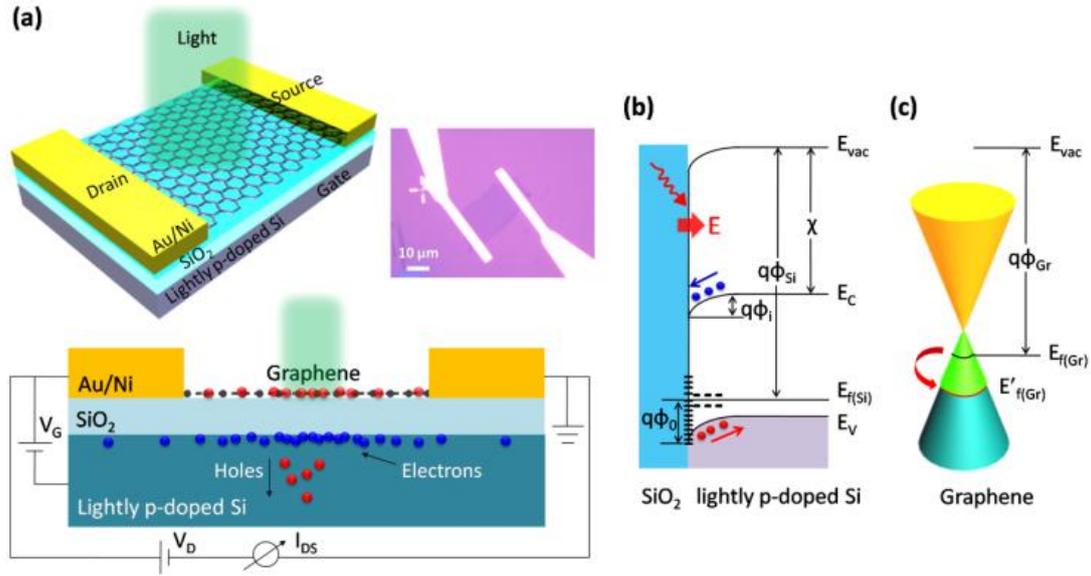

**Figure 1.** Graphene photodetector by interfacial gating. (a) Schematic diagram and optical image of the graphene photodetector on lightly p-doped silicon/SiO$_2$ substrate. (b,c) Schematic of energy band diagrams of the lightly p-doped silicon/SiO$_2$ interface with positive localized states (q$\varphi_0$) and its effect on graphene, respectively. The accumulation of photogenerated electrons (blue points) at the interface results in an additional negative voltage under light illumination, lowers the Fermi level ($E_{f(Gr)}$) to its new position ($E'_{f(Gr)}$) and thus a light-induced p-type doping in graphene.

## C. Photoresponse of graphene photodetectors

The photoresponse characteristics at $V_D = 1$ V and zero gate voltage ($V_G = 0$ V) are recorded with laser focused on the device at wavelength of 514 nm (spot size ~1 μm). Figure 2a shows the photoresponse of the channel current under different laser power, where positive photocurrents are observed when light is switched on. The dependence of the photocurrent as a function of light power is shown in Figure 2b. This photocurrent (at μA scales) is large enough for direct measurement without any amplifier, even at a very low light power (~0.6 nW). It should be noted that the photocurrent is saturated with the increase of light power. This is because the accumulation of photogenerated electrons at $SiO_2$/silicon interface will lead to a reversed electric field balance to the equilibrium built-in field. Correspondingly, less photo-induced electron/hole pairs will be separated as the net built-in field becomes weaker under higher illumination power. The responsivity of the device under different light power is calculated and shown in Figure 2c, which is defined as $R = I_{ph}/P$, where $I_{ph}$ and P are the photocurrent and incident light power, respectively. The device shows a remarkable responsivity up to ~1000 A $W^{-1}$ at an incident light power of ~0.6 nW, which is among the highest values of previously reported monolayer graphene photodetectors [1,2,5,21-24]. Based on this value of R, we also estimate external quantum efficiencies $EQE = R[h\upsilon/e] = 2.42 \times 10^5$ % and a specific detectivity $D^* = (A\Delta f)^{0.5} R/i_n = 1.1 \times 10^{10}$ Jones (1 Jones = 1 cm $Hz^{1/2}$ $W^{-1}$) at $V_D$= 1 V and $V_G$= 0 V, where e is electron charge, h is Planck's constant, $\upsilon$ is frequency of light, A is the effective area of the device, $\Delta f$ is the electrical bandwidth, R is the responsivity, and $i_n$ is the noise current (the dark current waveform of the device is shown in Supplement 1, Figure S2). Figure 2d shows the spectral photocurrent response of the device at ~0.05 μW light power from visible to near-infrared, which is obtained by a super continuum light source with a tunable filter. It can be seen that the excitation wavelength dependence of photocurrent is almost flat in visible regime and drops abruptly beyond ~1050 nm. This is a predictable outcome of the photoresponse mechanism proposed above, i.e. lightly-doped silicon is indeed the light absorbed medium, with photon response range from ~200-1100 nm. This also implies that by replacing silicon with other semiconductors, the photoresponse of device can be further extended to longer wavelength, e.g. HgCdTe with adjustable bandgap (from 0.7-25 μm) for mid-infrared photodetection [25].

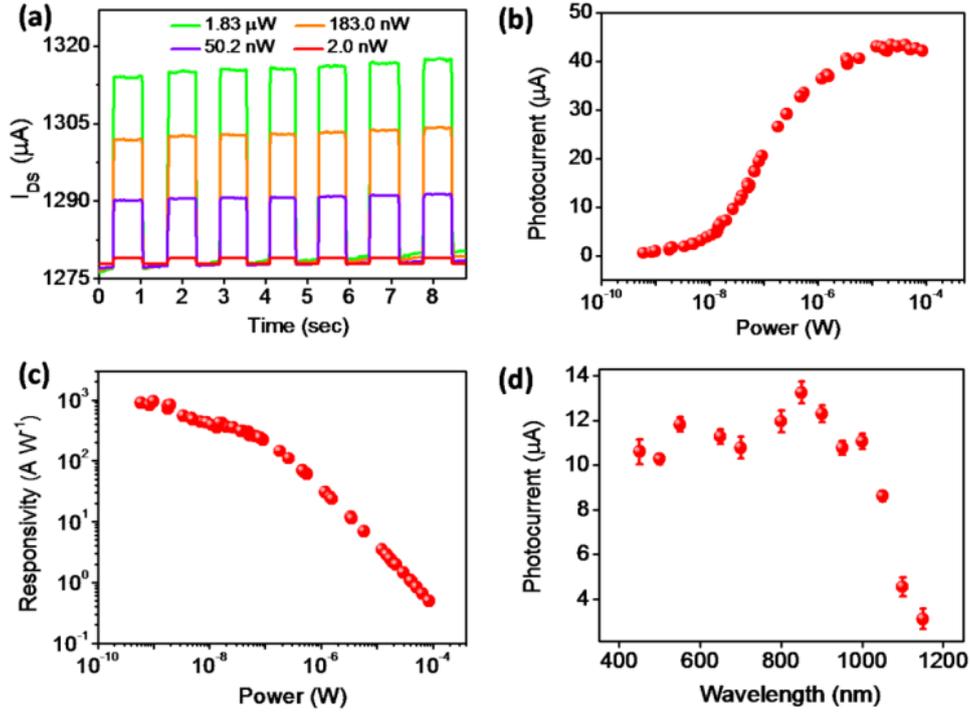

**Figure 2.** The photoresponse characteristics as a function of light power and wavelength. (a) Photo-switching characteristics of the graphene photodetector under different light power. (b,c) Photocurrent and responsivity at $V_D = 1$ V and $V_G = 0$ V of the device as a function of the light power. The laser wavelength is 514 nm. (d) The spectral photocurrent response of the device at ~0.05 μW light power from 450 to 1150 nm.

Next, we characterize the transfer characteristics of the device at $V_D = 10$ mV under different light power (with wavelength of 514 nm), as shown in Figure 3a. The transfer curves along with the Dirac points shift toward more positive gate voltage with the increase of light power. The inset is an enlarged view of the circled region in Figure 3a, showing the increase of light current with increased light power. This could be understood as a result from the additional negative gating effect caused by accumulated photogenerated electrons at the lightly-doped silicon/SiO$_2$ interface. In other words, a higher gate voltage is needed to obtain the charge neutrality point (Dirac point) in the graphene device. From these curves, we extract the shift of Dirac point ($\Delta V_G$) as a function of light power in Figure 3b (red points), which reaches saturated value of ~0.35 V at the power of ~10 μW. This corresponds to a modulated charge carrier density of $\Delta n = C_g \Delta V_G / e = $ ~$2.52 \times 10^{10}$ cm$^{-2}$, where $C_g$ is the dielectric capacitance ($1.15 \times 10^{-8}$ F cm$^{-2}$). Although the change of light-induced carrier density in graphene is less significant, the resulting photocurrent is rather considerable because of the high carrier mobility of graphene. The channel current change (ΔI) of the device under light illumination is defined by [11]:

$$\Delta I = I_{ph} = \frac{W}{L} C_g \mu \Delta V_G V_D,$$

where W and L are the width and the length of the graphene channel, respectively. According to this equation, the corresponding photocurrent is calculated to be ~0.41 μA, which is consistent with the measured photocurrent shown in Figure 2b (where the applied bias voltage $V_D$ is 100 times of here, i.e. 1V). Considering this, we also fabricate two devices of monolayer $MoS_2$ with low carrier mobility (0.1-10 $cm^2$ $V^{-1}$ $s^{-1}$ at room temperature [26]) on lightly and heavily p-doped silicon/$SiO_2$ substrates, respectively. The responsivities of these two devices show no obvious difference and are ~0.84 mA $W^{-1}$ at light power of ~1.5 μW (see Supplement 1, Figure S3). This again reveals the important role of high mobility graphene in the current device configuration. Figure 3c shows the corresponding photocurrents as a function of $V_G$ and are obtained by extracting the transfer curve in dark from those under illuminated conditions ($I_{light}$-$I_{dark}$). These curves look similar to the derivative of the transfer curve (dI/d$V_G$), indicating that the incident light can be treated as a gating field $\Delta V_G$, and the photocurrent $I_{ph} \propto \Delta V_G[dI/dV_G]$. Furthermore, at a fixed light power of 3.66 μW, we measure the photocurrents of the device at different gate voltages as plotted in Figure 3c (red circles), which are well consistent with the characteristic sigmoidal curve of photocurrent (blue line). It shows clearly that photoresponse can be reversed in sign and can even be switched off electrically by tuning the gate. Figure 3d shows the dependence of the photocurrent with different bias voltage $V_D$ under different light power. As expected, a linear dependence of the photocurrent is observed. The above results imply that the responsivity of our device can be effectively tuned, which is an attractive feature for developing tunable photodetectors for imaging applications, with responsivity adjustable to gate and bias voltages.

According to the photodetection mechanism described in Figure 1, the interfacial accumulated carriers will also diffuse in bulk Si in the lateral direction. To confirm this, we perform a spatial dependence of the photocurrent as a function of light position away from the graphene channel, as shown in Figure S4 (see Supplement 1). We find that the photocurrent still exists, even when the light spot is not on the graphene channel, whereas on the $SiO_2$ substrate, similar to photodetector based on graphene/Si junction [27]. The measured photocurrent is reduced when the light is illuminating away from the device under the same power. Based on the diffusion equation: $L_n = \sqrt{D_n \tau_n}$, where $L_n$ is the diffusion length of excess carriers, $D_n$ is diffusion coefficient, $\tau_n$ lifetime of carriers. In the case of lightly doped silicon with $N_A$= ~$10^{15}$ cm$^{-3}$, $\tau_n$= ~200 μs, $D_n$ = ~35 cm$^2$ s$^{-1}$, the calculated diffusion length $L_n$ is ~830 μm, which is consistent with the experimental results shown in Figure S4. On the other hand, we fabricate a control device with graphene lying on a heavily p-doped Si/$SiO_2$ substrate (resistivity ~$10^{-3}$ Ω cm). No obvious photocurrent could be resolved when the light (~1.8 μW) is switched on or off (see Supplement 1, Figure S5). The results clearly demonstrate that photocurrents in our device do not result from intrinsic photogenerated carriers in graphene; rather originate from the modulated charge carriers in the graphene due to the interfacial gating effect. More importantly, it is demonstrated that the lightly doped silicon/$SiO_2$ interface plays a critical role in the photosensitive behavior of our device. We have also studied the effects of dielectric thickness or materials (e.g. $Al_2O_3$) on the device performance (see Supplement 1, Figure S6 and S7). The results show that the thickness of dielectric layer has no obvious influences on the device performance because it will not affect the accumulated photoexcited charges at the interface. The photoresponse of the device on 30 nm $Al_2O_3$/lightly-doped Si substrate is also observed, while the responsivity is much lower compared to that of the devices on $SiO_2$/lightly-doped Si substrate.

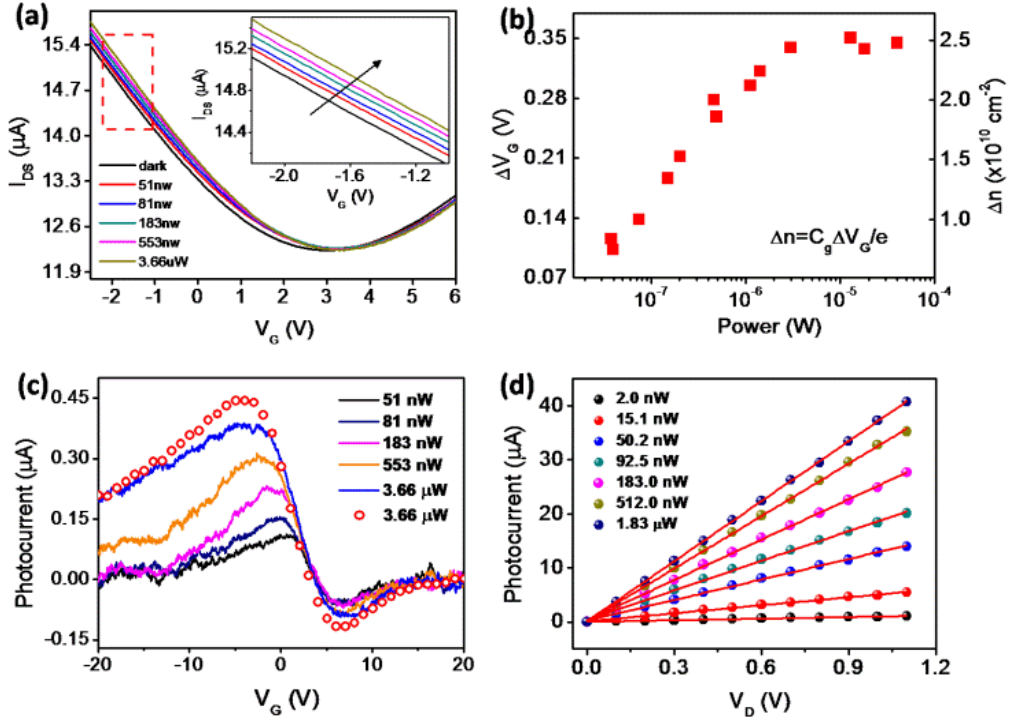

**Figure 3.** Gate and bias modulated photoresponse. (a) I-$V_G$ characteristics of the device under different light power, the inset is enlarged view of the circled region, showing the increase of light current under illumination with increased light power. (b) Horizontal shift of Dirac point ($\Delta V_G$) and the modulated charge carrier density ($\Delta n$) as a function of light power. (c) The extracted gate dependence of photocurrents ($I_{light}$-$I_{dark}$) from curves in (a). The red circles represent the photocurrents of the device at individual gate voltages under a fixed light power of 3.66 µW. (d) Photocurrents at $V_G = 0$ V of the device as a function of $V_D$ under different light power. The wavelength is 514 nm.

### D. Fast response time

Figure 4a shows the transient response of the device when light (514 nm, ~0.05 µW) is switched on or off by an acoustic optical modulator with frequency of 10 kHz. The rise ($\tau_{on}$) and fall ($\tau_{off}$) time are calculated to be ~400 and ~760 ns, respectively, based on curve fits of the transients with an exponential function. Such an ultra-fast response speed is superior to other graphene-based photoconductors with photo-gating mechanism [3,5-8,10-17]. More interestingly, the response time of our device increase very slowly with the decrease of light power, as shown in Figure 4b. On the other hand, although ultra-high responsivity has been achieved in graphene based hybrid structures and/or heterostructures, a significant increase of response time with the decrease of light power is commonly observed [24]. This behavior was also observed in phototransistors based on organic/inorganic composites before [28], and could be associated with the decay of transfer rate of electrons and/or holes from the light-absorbing materials to the conducting materials, especially in the case of weak light signal. The high speed response of our device is attributed to the fast separation of the electron-hole pairs assisted by the built-in electric field at the lightly doped silicon/$SiO_2$ interface. Specifically, the holes would be quickly driven into bulk Si before they recombine with the accumulated electrons, wherein there do not exist charge transfer process as common graphene based hybrid photodetectors do [3,5-17]. With the aim to further investigate the high speed photodetection of the device, we also perform the time-dependence of photoresponse at a high modulation frequency of 0.5 MHz under different light power, as shown in Figure 4c and 4d. It is demonstrated that our device could resolve weak signals at nWs level under high frequency operation, which is promising for high speed weak signal detections. Experimental results from additional device on 300 nm $SiO_2$/lightly-doped Si substrate are also shown in Figure S8 (see Supplement 1).

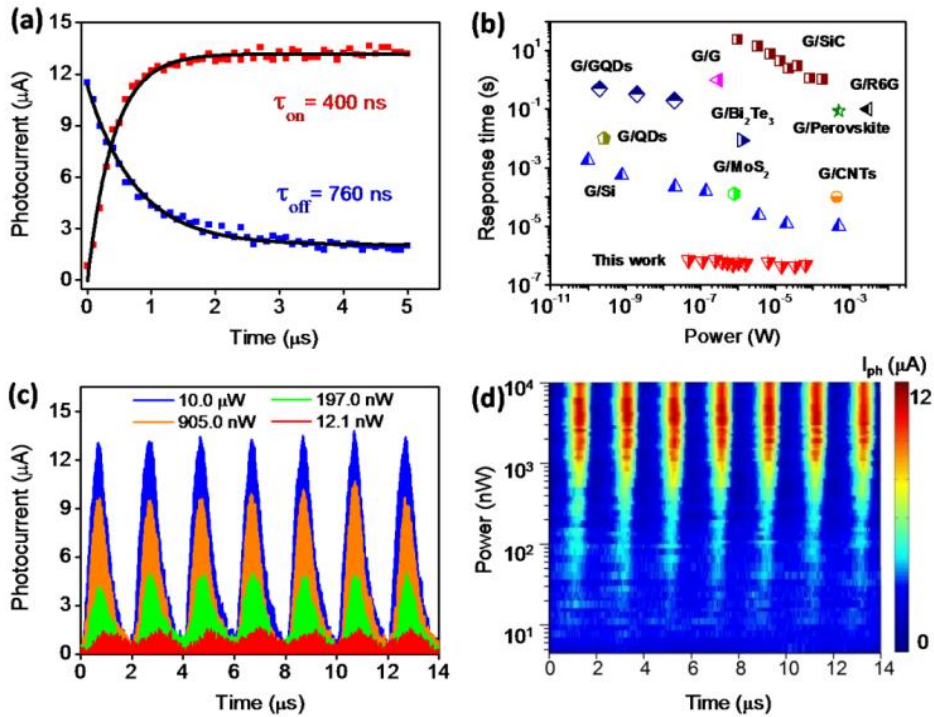

**Figure 4.** Transient response of the device. (a) The transient response of the device switched on or off by an acoustic optical modulator with frequency of 10 kHz. P= ~0.05 μW, $V_D$= 1 V

and $V_G$= 0 V. (b) The response time as a function of light power for our device and other graphene based photo-gating devices reported in the literature. (c,d) Photo-switching characteristics of the device at 0.5 MHz modulation frequency under different light power. The light wavelength is 514 nm.

## 4. CONCLUSION

In summary, by taking advantage of interfacial gating effect from lightly doped silicon/$SiO_2$ interface, we demonstrate a simple approach to graphene photodetection with high reponsivity and fast response. The proposed graphene photodetector exhibits high responsivity of ~1000 A $W^{-1}$ for weak signal of <1 nW and a spectral response that extends from visible to near-infrared. More importantly, the photoresponse time of our device has been pushed to ~400 ns and degrades quite slowly with the decrease of light power, which is superior compared to other graphene based photo-gating devices. Moreover, in comparison with the previous graphene-based devices with top gated p-n junctions [29,30], integrated with optical structures (e.g., plasmonic architecture [22], optical cavity [23], and waveguide [31]) and hybrid with light-absorbing materials (e.g., 2D vdW crystals [4,7,10,14,15], QDs [3,8,11], nanowire/tube [6,9]), our device possesses the advantages of simple fabrication process and is fully compatible with the silicon technology. This work therefore not only opens up a route to graphene-based high performance optoelectronic devices, but also provides the potential to access an even wider spectral range by combing graphene with other oxide-semiconductor system.

**Funding.** National Natural Science Foundation of China (NSFC) (61422503, 61376104); The open research funds of Key Laboratory of MEMS of Ministry of Education (SEU, China); The Fundamental Research Funds for the Central Universities; The open research fund of SEU-JGRI Joint Research Center of Advanced Carbon Materials.

See Supplement 1 for supporting content.

responsivity," Nat. Photon. **7**, 883–887 (2013).

# Supplement 1

1. **Characterizations of graphene device**

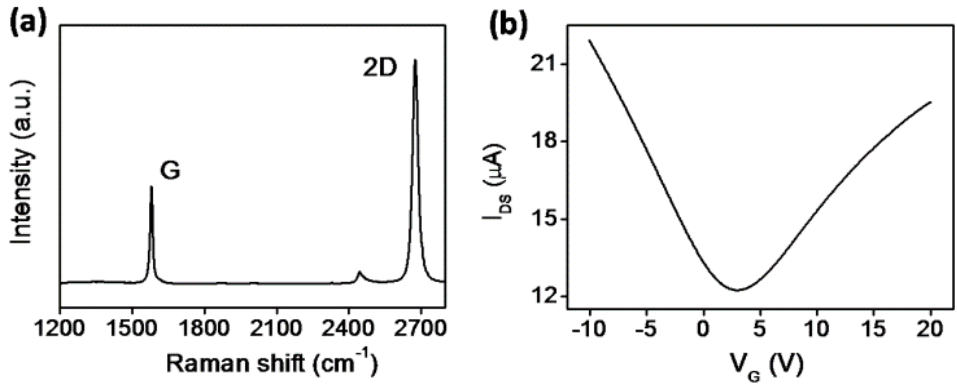

**Figure S1.** (a) Raman spectrum of graphene deposited on 300 nm SiO$_2$/lightly p-doped silicon substrate. G and 2D bands locate at 1580.3 and 2675.2 cm$^{-1}$, respectively. I$_{2D}$/I$_G$=~2.29, and the full width at half maximum of 2D band is 27.7 cm$^{-1}$. (b) A transfer characteristic at an applied bias voltage V$_D$=10 mV of the device measured in dark.

2. **Dark current waveform of graphene device**

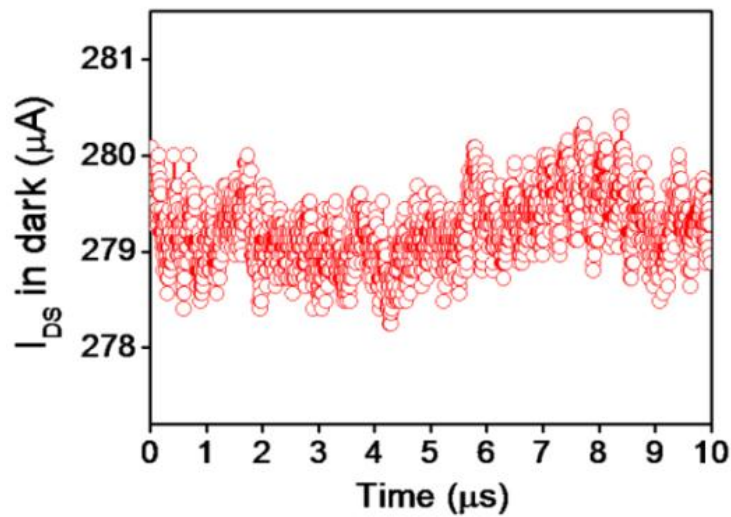

**Figure S2.** The dark current waveform of the graphene photodetector measured at an applied bias voltage of 1 V. The electrical bandwidth Δf= 2.5 ×10$^8$ Hz, the noise current i$_n$=~1.0 μA.

## 3. MoS$_2$ devices

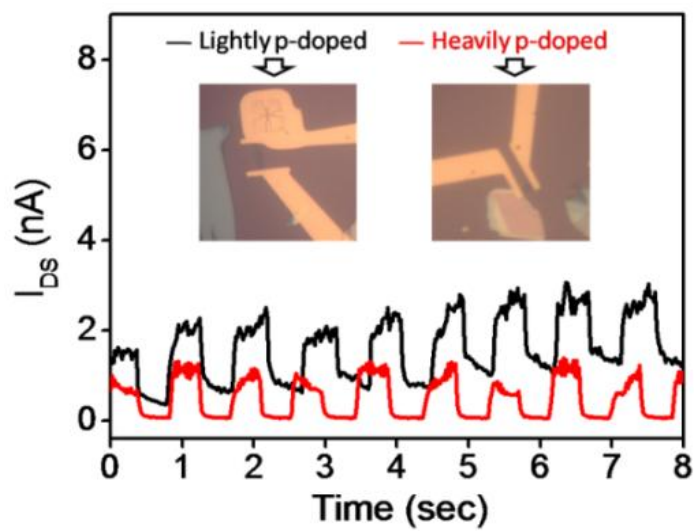

**Figure S3**. The time-dependence of photoresponse of MoS$_2$ devices under light illumination ($\lambda$=514 nm, P =1.5 μW, $V_D$=1 V, $V_G$=0 V), the insets show the optical images of monolayer MoS$_2$ devices on lightly and heavily p-doped silicon/SiO$_2$ substrates, respectively.

## 4. Spatial dependence of photocurrent

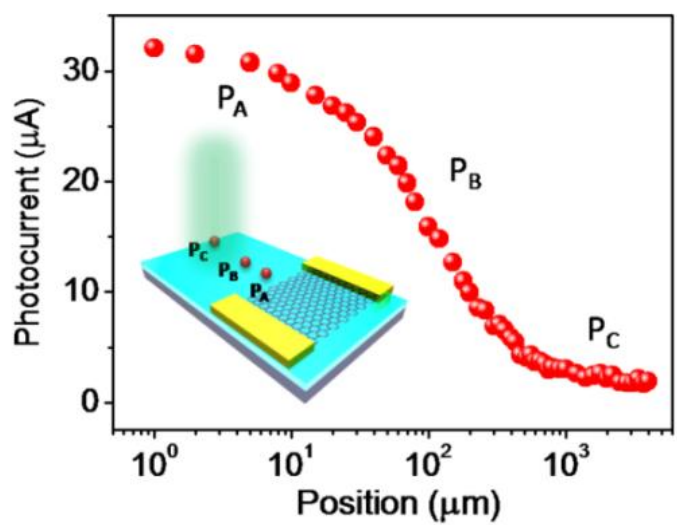

**Figure S4.** The spatial dependence of the photocurrent as a function of light illumination position away from the graphene channel on ~300 nm SiO$_2$/lightly-doped Si substrate.

## 5. Device on 300 nm SiO$_2$/heavily p-doped Si

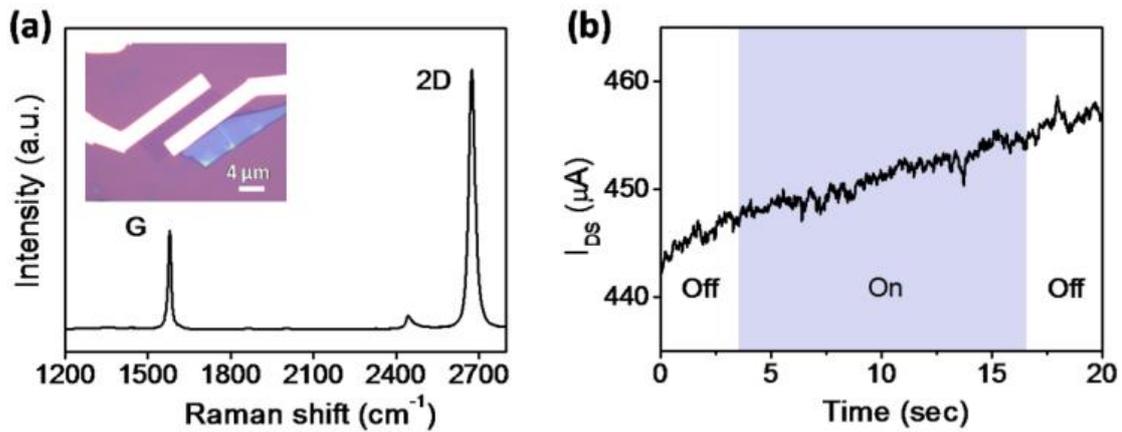

**Figure S5.** (a) Raman spectrum and optical image of the control graphene device on a heavily p-doped silicon/SiO$_2$ substrate (~10$^{-3}$ Ω cm). G and 2D bands at ~1579.1 cm$^{-1}$, ~2674.7 cm$^{-1}$, respectively, I$_{2D}$/I$_G$=~2.63, the full width at half maximum of 2D band is ~26.9 cm$^{-1}$. (b) No obvious photocurrent could be resolved within the measurement resolution of the electronics when switching on and off the light (λ= 514 nm, P =1.8 μW, V$_D$=1 V).

6. **Devices on SiO$_2$/lightly-doped Si substrates with different SiO$_2$ thicknesses**

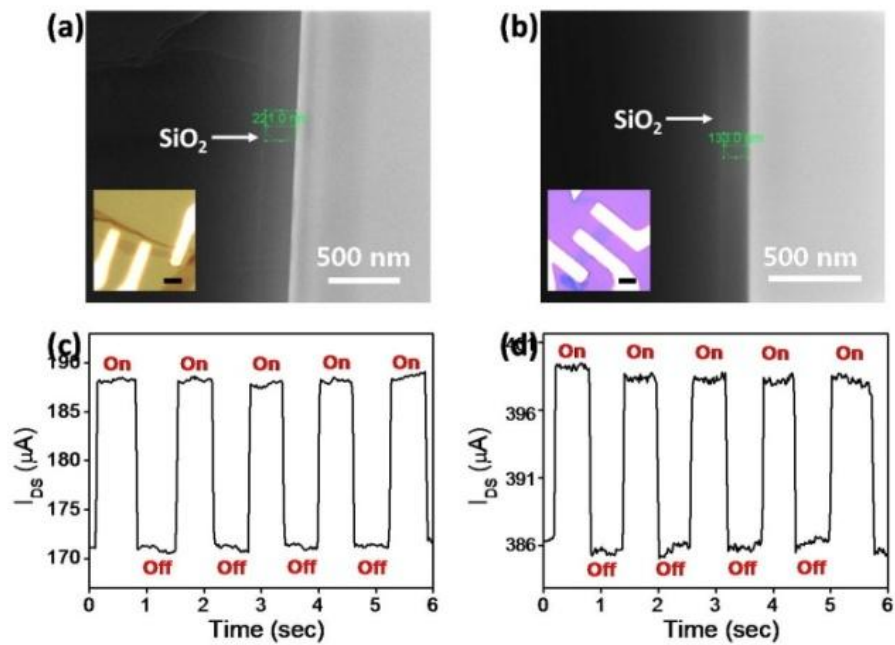

**Figure S6.** (a,b) SEM images of the cross section of SiO$_2$/lightly-doped Si substrates with different SiO$_2$ thicknesses. The insets show the optical images of the corresponding graphene devices. Scale bar=5 μm. (c,d) Time-dependent photocurrent of the graphene devices fabricated on ~220 nm and ~130 nm SiO$_2$/lightly-doped Si substrates. The average

photocurrent is ~17.0 μA and ~13.5 μA at power of ~0.06 μW, respectively, which is comparable to that of the devices on ~300 nm SiO$_2$/lightly-doped Si substrates (~17.3 μA) in the main text, suggesting that the thickness of dielectric layer has no obvious influence on the device performance. ($\lambda$= 514 nm, V$_D$=1 V).

## 7. Device on 30 nm Al$_2$O$_3$/lightly-doped Si

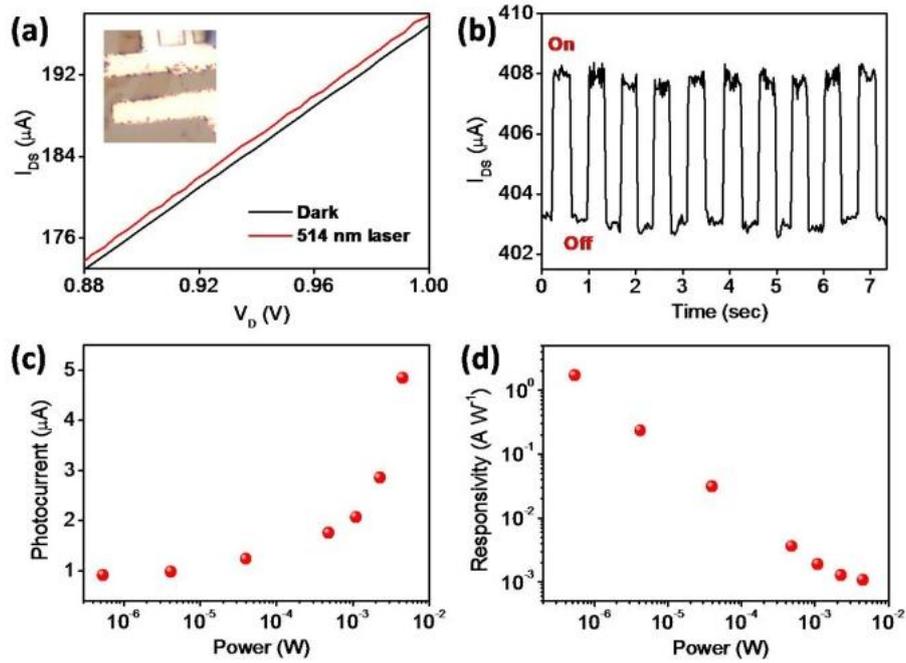

**Figure S7.** Current-voltage characteristics of graphene device on 30 nm Al$_2$O$_3$/lightly-doped Si substrate under dark and illuminated conditions, the inset shows the optical image of the device. (b) A typical dynamic photocurrent response of the device at power of ~4.5 mW. (c,d) Photocurrent and responsivity at V$_D$ =1 V and V$_G$ =0 V of the device as a function of the light power. The responsivity of this device is much lower compared to that of the devices on SiO$_2$/lightly-doped Si substrate.

## 8. Experimental results from additional device on 300 nm SiO$_2$/lightly-doped Si substrate

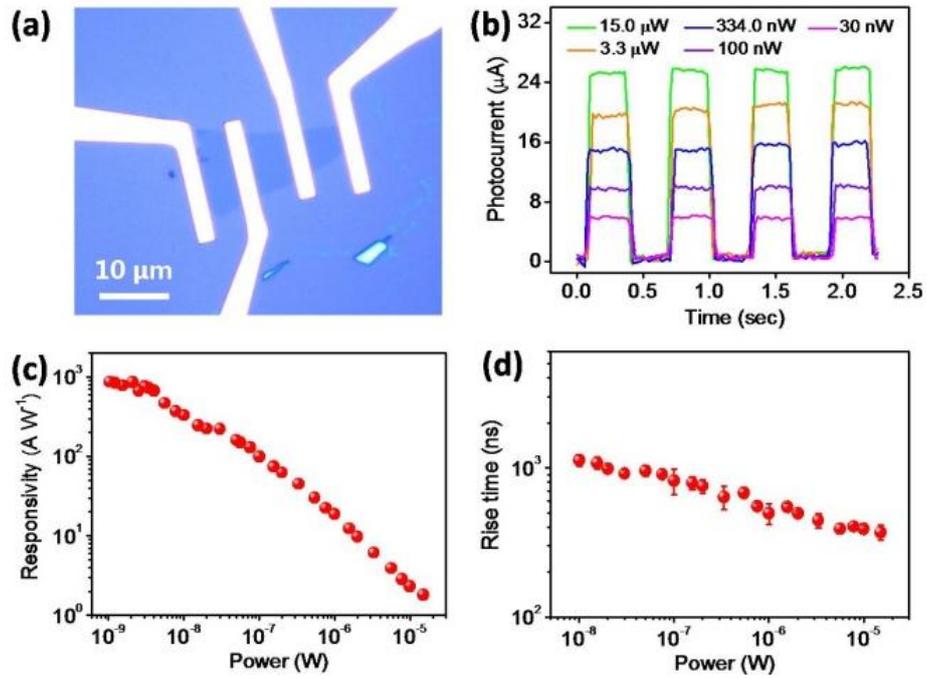

**Figure S8.** (a) Optical image of another graphene device on 300 nm $SiO_2$/lightly-doped Si substrate. (b) The dynamic photocurrent response of the device under different light power. (c,d) Responsivity and response time as a function of light power under the same experimental conditions as the device shown in the main text.